\documentclass[runningheads]{llncs}
\usepackage[T1]{fontenc}
\usepackage{graphicx}
\usepackage{amsmath}
\usepackage{booktabs,multirow}
\usepackage{xcolor}
\usepackage{booktabs}
\usepackage{multirow}
\usepackage{geometry}
\usepackage{array} 
\usepackage{makecell}
\usepackage{algorithm}    
\usepackage{algpseudocode}
\usepackage{float}        
\usepackage{amssymb}
\DeclareMathOperator*{\argmax}{arg\,max}

\algdef{SE}[CLASS]{Class}{EndClass}[1]{\textbf{class} #1}{\textbf{end class}}

\begin{document}
\title{AlphaSeek: Trajectory-Level Self-Iterative Factor Mining Framework for Multi-Source Financial Data}
\titlerunning{AlphaSeek}
\author{
Qilu Zhu\inst{1}\orcidID{0009-0008-1687-3814} \and
Zijun Lu\inst{1}\orcidID{0009-0002-6305-3065} \and
Jianmin Zhu\inst{1}\orcidID{0009-0001-9597-1533} \and
Ning Chen\inst{1}\orcidID{0000-0001-6607-1758}\thanks{Corresponding author.} \and
Shuo Yin\inst{2}\orcidID{0009-0004-5298-5822}  \and  \\
Simon Fong\inst{3}\orcidID{0000-0002-1848-7246}
}
\authorrunning{Q. Zhu et al.}
\institute{
Zhongnan University of Economics and Law, Wuhan 430073, China\\
\email{
qiluzhu@stu.zuel.edu.cn,
lzj0489@gmail.com,
superju454@gamil.com,
safetychn@zuel.edu.cn
}
\and
Tsinghua University, Beijing 100084, China\\
\email{yins25@mails.tsinghua.edu.cn}
\and
University of Macau, Macau 999078, China\\
\email{ccfong@um.edu.mo}
}
\maketitle
\begin{abstract}
With the rapid rise of large language models, LLM-driven quantitative factor mining has become an increasingly active research area due to its potential to unify financial knowledge understanding and automated alpha discovery. However, existing methods still suffer from subjective direction design, limited integration of up-to-date multi-source information, semantic drift, factor redundancy, and the absence of an end-to-end feedback loop from factor discovery to portfolio backtesting. To address these limitations, we propose AlphaSeek, an end-to-end factor mining framework for quantitative investment that integrates automated direction discovery, trajectory-level factor evolution mining and self-iterative portfolio optimization. AlphaSeek first automatically collects and summarizes multi-source financial information, including quantitative papers, news et al. to identify promising mining directions. It then performs trajectory-level factor mining by extending the optimization unit from a single factor expression to a complete research trajectory covering hypothesis generation, factor construction, validation, backtesting, and feedback. Based on this design, we introduce evolution operators—parallel direction expansion, mutation and crossover—to improve search diversity, refinement quality and factor robustness. Finally, AlphaSeek constructs a self-iterative factor portfolio, allowing newly discovered factors to interact with an existing state-of-the-art(SOTA) factor library under redundancy-aware constraints. Experiments on CSI300 show that AlphaSeek achieves the strongest overall strategy-level performance on CSI300 with ARR of \textbf{8.28\%}, IR of \textbf{1.29\%} and MDD of \textbf{6.28\%}, while remaining competitive on factor predictive metrics with IC of \textbf{0.0454} like Fig. \ref{Model Comparison}, while Fig. \ref{fig:csi500} shows factors mined on CSI300 also achieve strong time-series return performance on CSI500 than other models, suggesting promising cross-market transferability under a zero-shot setting.

\end{abstract}

\begin{figure}[t]
  \centering
  \begin{minipage}{0.40\linewidth}
    \centering
    \includegraphics[width=\linewidth]{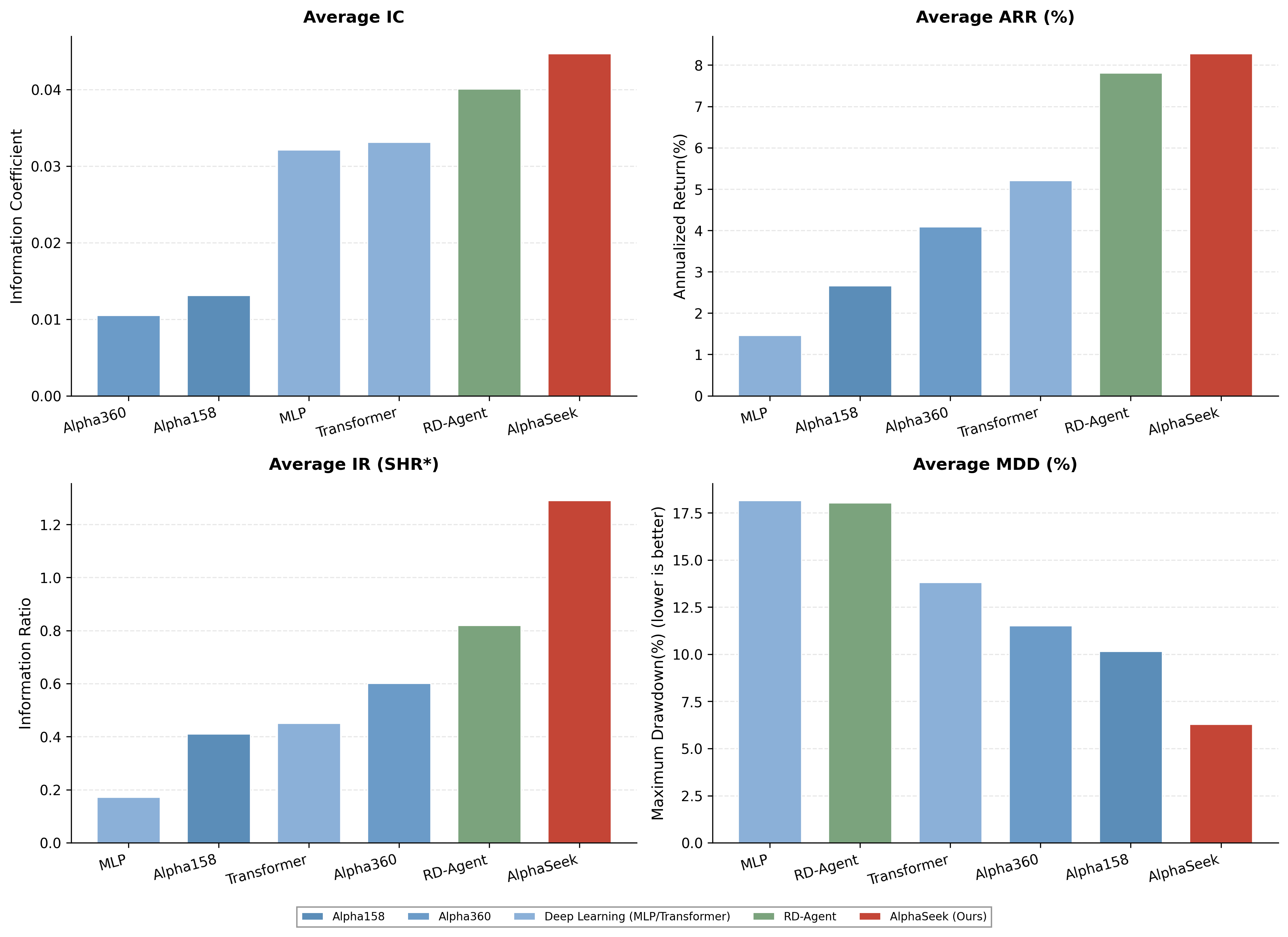}
    \caption{Model Comparison on CSI300}
    \label{Model Comparison}
  \end{minipage}
  \hfill
  \begin{minipage}{0.48\linewidth}
    \centering
    \includegraphics[width=\linewidth]{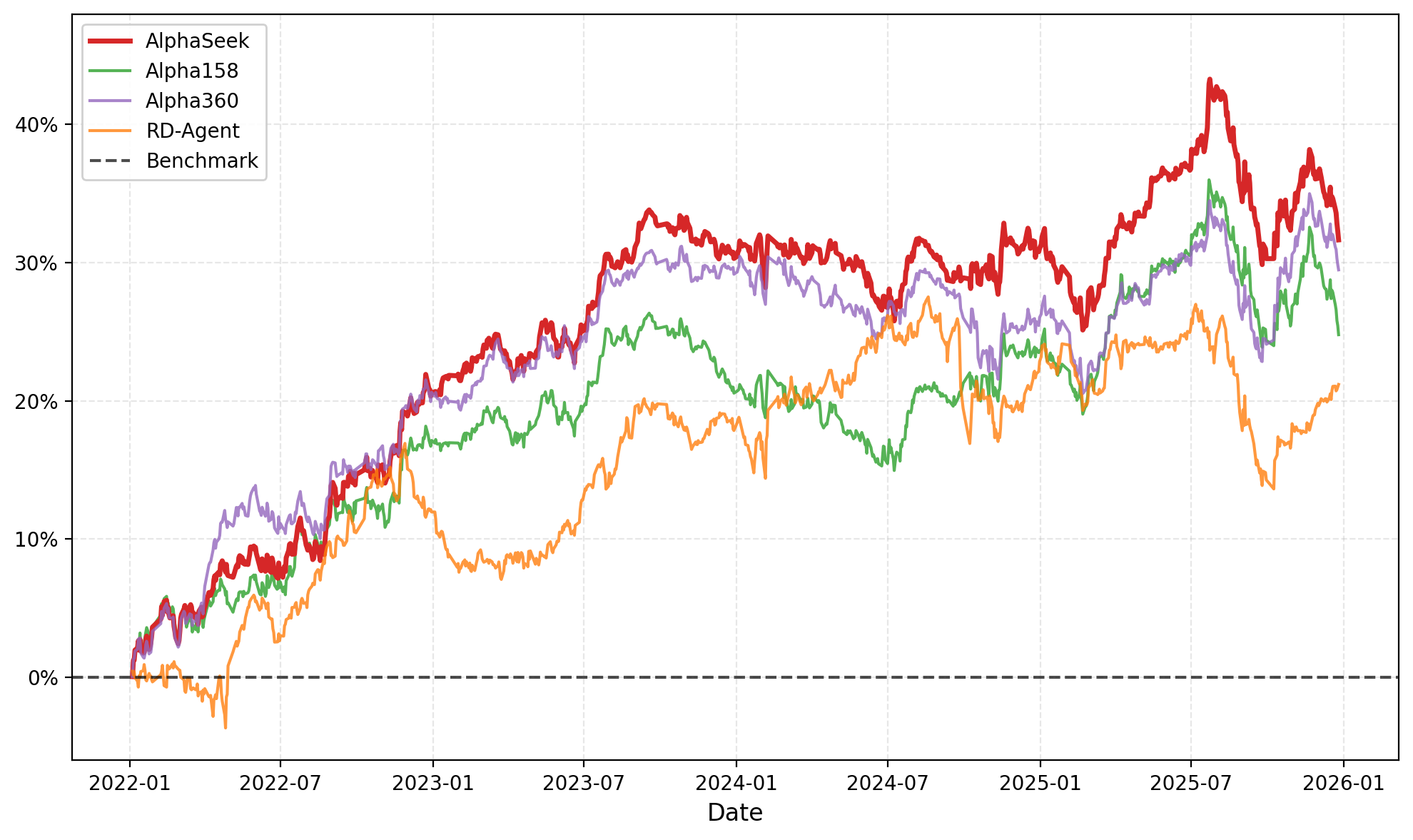}
    \caption{Time-Series Returns on CSI500}
    \label{fig:csi500}
  \end{minipage}
\end{figure}

\section{Introduction}

Quantitative factor mining is one of the core issues in modern quantitative investment, whose development has undergone a long-term theoretical and technological evolution. The earliest equilibrium explanation of risk and return can be traced back to the Capital Asset Pricing Model (CAPM) \cite{sharpe1964,lintner1965}. Subsequently, the Fama–French Three-Factor Model introduced market, size and value factors into the asset pricing system \cite{fama1993,carhart1997}, and was further extended to the Five-Factor Model including profitability and investment factors \cite{fama2015,fama2016,ang2006}. On this basis, a large number of empirical studies have explored the impact of macroeconomic, behavioral, liquidity and other factors on stock returns \cite{novy-marx2013,harvey2016,demiguel2009}.

To realize automatic factor search, Genetic Programming (GP) \cite{yang2017sentimentgp,kamijo1990} and Reinforcement Learning (RL) \cite{yang2020drlt,moody1998,li2018} have been widely applied to factor construction and strategy generation. Deep learning (DL) techniques \cite{ma2025evoir,hu2025clusir,hu2026proto,hu2026spikerestormer} have also been used to extract nonlinear factor representations from high-dimensional market data \cite{heaton2017,fischer2018,bao2017}. In recent years, LLM-driven factor mining methods have developed rapidly, which expand the candidate factor space through natural language understanding capabilities and improve factor quality combined with structured search mechanisms \cite{tang2025alphaagent,han2026quantaalpha,shi2025alphajungle,li2023llmfinance}.

However, whether traditional manual construction or automated search frameworks, mostly rely on the subjective experience of researchers or static fixed datasets \cite{harvey2016}. Classic studies have confirmed that the profitability of pricing factors will decay significantly after academic publication \cite{mclean2016}, which makes it critical to capture cutting-edge asset pricing achievements in a timely manner. What is frustrating is traditional paradigms can hardly adapt to dynamic market changes and update research directions synchronously with the latest academic progress. More importantly, this critical gap has not been fully addressed in mainstream LLM-driven factor mining frameworks \cite{tang2025alphaagent,han2026quantaalpha}, and few studies have established an automated mechanism to extract potential factor mining directions from up-to-date academic literature.

Nevertheless, LLM-driven agents still face multiple core challenges in factor mining: the generated factor expressions are prone to semantic drift and overfitting to historical data, along with insufficient multi-source information integration capabilities \cite{han2026quantaalpha,tang2025alphaagent,han2026,shi2025alphajungle,li2023llmfinance,wang2024}. Furthermore, most current factor mining directions rely on expert-designed paradigms with strong subjectivity, leading to high redundancy in the mined factors and limiting the diversity of the candidate factor space. Moreover, portfolio backtest is a shortcoming as most research focuses on backtesting individual factors. Existing methods attempt to solve partial problems through self-training or multi-agent collaborative frameworks \cite{tang2025alphaagent,han2026quantaalpha,rdagentq2025,han2026}, but it is difficult to balance the diversity, robustness and interpretability of factors simultaneously. 

To address the above limitations, we proposed AlphaSeek, an end-to-end automated quantitative information mining agent, with core contributions as follows:

\begin{itemize}
\renewcommand{\labelitemi}{\textbullet} 
    
    \item  \textbf{An automated mining direction acquisition mechanism} is designed, which can automatically retrieve and summarize the latest quantitative finance academic papers, extract potential innovative factor research directions.

    \item \textbf{A trajectory-level factor mining framework} is proposed, which uses trajectory as the optimization unit containing the complete research life cycle.

    \item \textbf{An end-to-end automated closed loop} of "direction exploration - factor mining - portfolio backtest " is constructed, and dynamic self-interaction optimization between new factors and the SOTA factor library is realized.

\end{itemize}

\begin{figure}[t]
  \centering
  \includegraphics[width=0.95\linewidth]{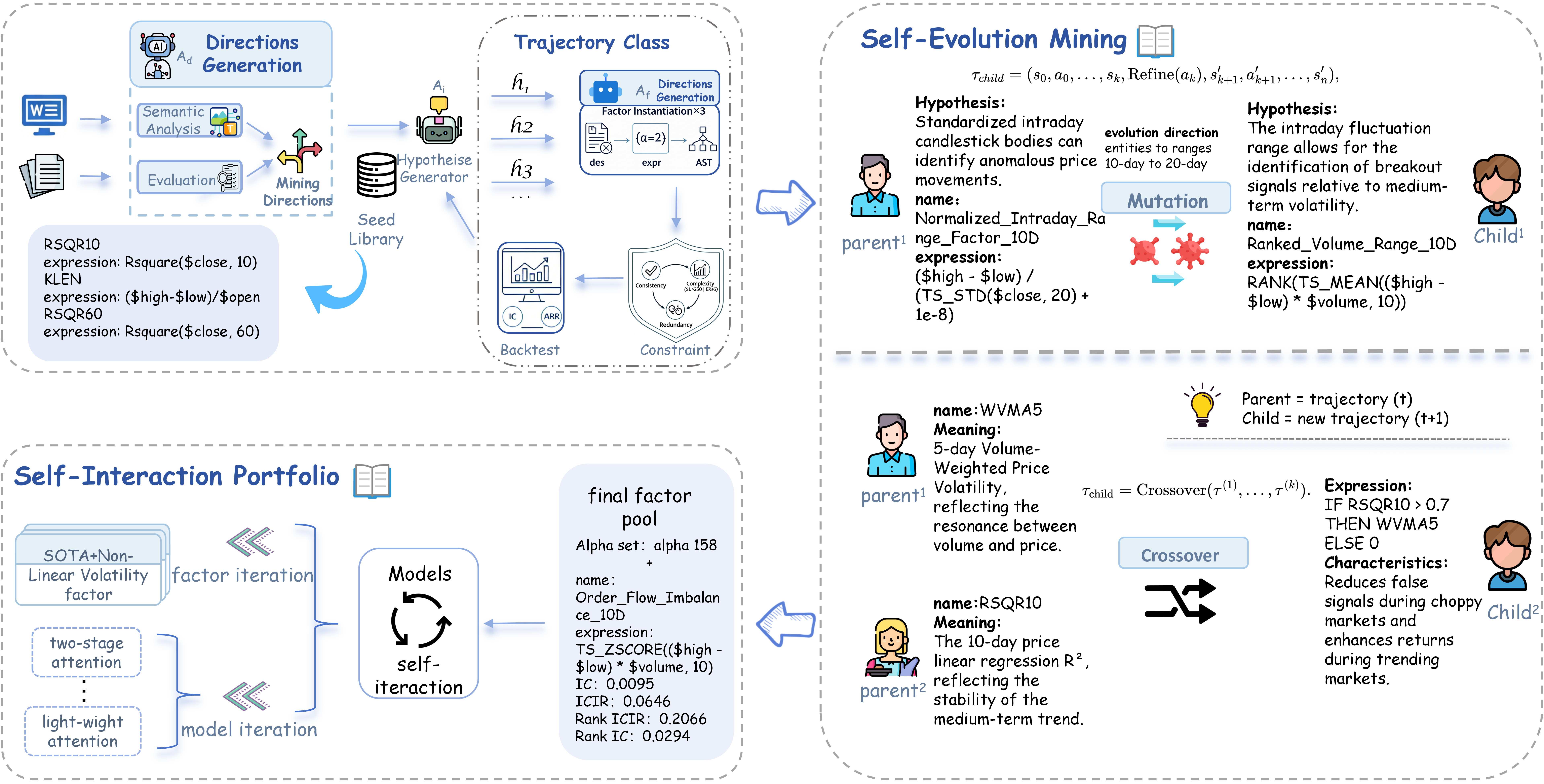}
  \caption{Overall Architecture of AlphaSeek: An End-to-End Quantitative Mining Agent: From Direction Generation to Single-Factor Trajectory Evolution Mining, and Finally to Self-Interactive Portfolio Construction}
  \label{fig:system_overview}
\end{figure}

\section{Related Work}

\subsection{Classic Quantitative Factor Mining}
 While the classic linear factor pricing framework represented by CAPM and Fama-French series models has laid the core theoretical foundation for modern factor investing \cite{sharpe1964,fama1993,fama2015}, subsequent empirical studies have gradually exposed two inherent bottlenecks of traditional factor mining paradigms: the proliferation of published factors has brought severe data snooping and overfitting risks systematically verified by Harvey et al. \cite{harvey2016}, and manual factor construction heavily relies on expert experience, which cannot efficiently explore the high-dimensional nonlinear feature space in massive market data. To break through these limitations, automated factor search technologies have undergone long-term development: GP-based methods realize automatic traversal of factor expressions through evolutionary algorithms, but they are essentially random searches without economic logic constraints, resulting in poor interpretability and generalization ability of generated factors \cite{yang2017sentimentgp}; RL-based methods optimize factor screening with long-term return objectives, but they are prone to fall into local optima in high-dimensional feature space and lack effective overfitting control mechanisms \cite{yang2020drlt,li2018}; DL technologies can extract complex nonlinear factor representations, but they face an intractable "black box" problem, where the generated factors cannot be traced back to clear economic logic and are difficult to adapt to the compliance requirements of actual investment \cite{heaton2017,gu2020}. All the above traditional automated methods take single factor expressions or strategy weights as the core optimization unit, lacking a complete record of the full life cycle of factor research and knowledge inheritance, and fail to establish a full-process constraint mechanism for the factor mining process, making it difficult to fundamentally solve the problems of overfitting and insufficient generalization ability, which are exactly the core issues addressed by the trajectory-level factor optimization framework in this paper.

\subsection{Portfolio and Backtesting Optimization}
 Multi-factor portfolio construction and backtesting are the core links for translating factor mining achievements from theoretical research to live trading implementation, with the core objective of enhancing portfolio robustness through multi-factor fusion while controlling transaction costs and overfitting risks. In terms of factor combination methods, the classic research by DeMiguel et al. verified the robustness of equal-weighted portfolios in most market scenarios and provided a fundamental benchmark for multi-factor portfolio construction \cite{demiguel2009}, and subsequent studies have gradually developed optimization methods such as risk parity, maximum Sharpe ratio, minimum variance, and dynamic factor combination strategies with time-varying weights to further improve the adaptability of multi-factor portfolios in complex market environments \cite{leung2019,wang2024,zhou2025}; In the field of backtesting optimization, information coefficient (IC)-based factor similarity measurement has become the mainstream method, which can effectively eliminate highly correlated redundant factors in the factor library to reduce the overfitting risk and crowding degree of the portfolio \cite{rdagentq2025}, yet most existing studies adopt a batch backtesting mode based on static factor pools, lacking an incremental factor fusion mechanism and failing to realize the dynamic collaborative optimization of newly mined factors and the existing SOTA factor library, which is the key problem solved by AlphaSeek.

\subsection{Agent-Based Alpha Mining}
 The application of LLM in quantitative finance has rapidly evolved from early unstructured text processing for sentiment and fundamental information extraction \cite{tang2025,chen2025,liu2024}, to code and formula generators that replace or supplement traditional GP methods for automated Alpha factor construction, represented by the AlphaFin framework with retrieval-augmented generation fine-tuning and prompt engineering-based trading formula iterative evolution methods \cite{li2024alphafin,li2024fama}, and the latest research frontier has focused on integrating LLMs' semantic understanding, code generation and logical reasoning capabilities into agent systems to realize adaptive iteration in dynamic market environments \cite{han2026}, including the QuantAgent system with self-training mechanism, the interactive Alpha-GPT paradigm, and the closed-loop alpha mining framework of AlphaAgent \cite{wang2024quantagent,liu2023alphagpt,tang2025alphaagent,han2026quantaalpha}. Despite these advances, existing LLM-driven factor mining methods still have core limitations: they face a high risk of semantic drift in factor generation, lack full-process risk constraints for the mining pipeline, and cannot balance the diversity, robustness and interpretability of factors simultaneously.

\section{Method}\label{sec3}

\subsection{Problem Setup}
\subsubsection{Alpha Mining} The factor mining task aims to learn effective alpha factors $f$ from the market feature tensor $\mathbf{X} \in \mathbb{R}^{N \times T \times D}$ corresponding to the stock pool $\mathcal{S} = \{s_1, \dots, s_N\}$ and time interval $\mathcal{T} = \{t_1, \dots, t_T\}$. On each trading day $t \in \mathcal{T}$, the factor generates prediction signals from the feature slice $\mathbf{X}_t \in \mathbb{R}^{N \times D}$ to forecast the cross-sectional stock returns $\mathbf{y}_{t+1} \in \mathbb{R}^N$. Formally, an Alpha factor can be expressed as a mapping $f(\mathbf{X}_t) \to \mathbf{y}_{t+1}$, where $\mathbf{y}_{t+1}$ is the ground truth return in period $t+1$. Accordingly, this paper formulates Alpha factor mining as the following optimization problem:
\begin{equation}
f^* = \argmax_{f \in \mathcal{F}} \mathcal{L}(f(\mathbf{X}), \mathbf{y}) - \lambda \mathcal{R}(f),
\end{equation}
where $\mathcal{F}$ denotes the space of all feasible factor expressions, $\mathcal{L}(\cdot)$ measures the predictive effectiveness of the factor, which is quantified by the time-series average information coefficient $\text{IC}(f)$ of the factor \cite{li2024fama,duan2022factorvae}. $\mathcal{R}(\cdot)$ is the regularization term, which integrates complexity penalty, redundancy penalty and consistency penalty to constrain the conciseness, novelty and live trading applicability of the factor expression \cite{tang2025alphaagent}. $\lambda$ is the weight coefficient used to balance the predictive profitability of the factor and the strength of regularization constraints.

\subsubsection{Alpha Mining Trajectory} Different from purely data-driven factor mining pipelines, AlphaSeek introduces market investment hypotheses $h \in \mathcal{H}$ to guide the LLM-based factor construction process. Each complete round of alpha mining follows an end-to-end workflow from hypothesis generation, factor construction to backtesting evaluation. AlphaSeek defines each complete mining process as a mining trajectory, formalized as an ordered sequence $\tau = (s_0, a_0, s_1, a_1, \dots, s_n)$, where $s_0$ is the initial mining context, $a_i$ is the action executed by the multi-agent system at the $i$-th step, and $s_n$ is the terminal state containing the evaluation results of this round of mining. The quality of a trajectory is measured by its terminal reward \cite{han2026quantaalpha,yu2023synergistic}, formally defined as:
\begin{equation}
R(\tau) = \mathcal{L}(f_\tau(\mathbf{X}), \mathbf{y}) - \lambda \mathcal{R}(f_\tau),
\end{equation}
where $f_\tau$ denotes the alpha factor finally generated by the trajectory $\tau$, and $\tau \in \mathcal{T}$, $\mathcal{T}$ denotes the complete strategy trajectory space).

\subsubsection{Objective} The core optimization objective of AlphaSeek is to solve the optimal trajectory generation policy $\pi$, to maximize the expected terminal reward of the generated mining trajectories \cite{han2026quantaalpha,yu2023synergistic,tang2025alphaagent}, formally expressed as:
\begin{equation}
\pi^* = \argmax_{\pi} \mathbb{E}_{\tau \sim \pi} \left[ R(\tau) \right],
\end{equation}
where $\tau \sim \pi$ denotes trajectory samples obtained by repeatedly applying the trajectory generation policy $\pi$ starting from the initial state $s_0$.

\subsection{AlphaSeek}
The overall architecture of AlphaSeek see Fig. \ref{fig:system_overview} includes three core components: direction generation, single-factor trajectory strategy mining and incremental portfolio strategy, which fully realizes a fully automated end-to-end closed loop from row data to hypothesis input and then to strategy performance output. The core logic of the system is as follows: raw data derived from quantitative research reports is processed and summarized by $A_d$ to generate directions for factor mining, and then multiple complementary exploration hypothesis proposed by $A_i$ enter a five-step cycle of "hypothesis generation → factor construction → factor calculation → backtesting verification → feedback summary", which is executed in each direction to form a strategy trajectory that records the complete life cycle of factor research. Subsequently, the trajectory is continuously optimized through the evolutionary cycle of Original→Mutation→Crossover, and the cycle repeats until convergence, so as to mine effective factors with sound economic logic and predictive power. Finally, the effective factors that pass the constraint verification are continuously identified and integrated into the SOTA factor portfolio, and the continuous improvement of strategy returns is realized through incremental portfolio optimization, forming a complete iterative closed loop of "mining-verification-feedback-optimization". 

\begin{algorithm}[H]
\caption{Definition of the Trajectory Class}
\begin{algorithmic}[1]
\Class{Trajectory}
    \State \quad $Hypothesis\ h$ \Comment{Investment hypothesis text corresponding to the factor}
    \State \quad $Factor\ f \in \mathcal{F}$ \Comment{Set of factor expressions generated by the trajectory}
    \State \quad $Code\ $ \Comment{Executable code implementation of the factor}
    \State \quad $Backtest\ Result\ $ \Comment{Set of single-factor backtesting results}
    \State \quad $Evaluation\ Metrics\ $ \Comment{Evaluation metrics of factor predictive power}
    \State \quad $Feedback\ \phi$ \Comment{Feedback summary for iterative optimization}
    \label{Trajectory Class}
\EndClass
\end{algorithmic}
\end{algorithm}

\subsubsection{Direction Generation}
We focuses on the automated access to various core data such as news information and academic research. After preprocessing the multi-source heterogeneous data, in-depth semantic analysis is carried out: firstly, semantic encoding is performed on news headlines, abstracts or core viewpoints of papers to extract key financial entities, event types and internal logical relationships. Subsequently, the matching degree between the information and the target market is evaluated, and only when the matching degree exceeds the preset access threshold, the relevant information is included in the candidate set for further analysis. For the highly relevant information entering the candidate set, qlpha extracts it and outputs a set of factor mining direction sets:
\begin{equation}
\small
\mathcal{S}_{\text{dir}} = \text{LLM} \left( \bigcup_{k \in \{\text{News, Papers}\}} \text{Preprocess}(D_k) \ \middle|\ \text{Constraint: Datasets} \right)
\end{equation}

\subsubsection{Trajectory Mining}
Given an input natural language research direction $s \in \mathcal{S}_{\text{dir}}$, the factor generation process is completed with standardized symbolic constraints: first, the Idea Agent $\mathcal{A}_i$ maps the research direction $s$ to a formalized parallel investment hypothesis $h \in \mathcal{H}$, then the Factor Agent $\mathcal{A}_f$ that receives the hypothesis $h$ generates the corresponding symbolic factor expression $f \in \mathcal{F}$ based on the standardized operator library $\mathcal{O}$ and raw market feature space $\mathcal{X}$(like CSI300), parses $f$ into an Abstract Syntax Tree (AST) representation $T(f)$ for structural validity verification, and compiles $T(f)$ into executable code $c$ to complete the fully controllable end-to-end construction of factors. Different from traditional methods that take a single factor expression $f$ as the core optimization unit, this module innovatively proposes a trajectory-based factor mining paradigm $\tau$, which upgrades the optimization unit from the volatile "single factor" to the "strategy trajectory" covering the full research life cycle of the factor, realizing the traceable inheritance of factor research knowledge and full-link interpretability of the entire mining process. The structured definition of the strategy trajectory is standardized via pseudo-code as Algorithm 1.

\subsubsection{Constraint Control}
We consider complexity, redundancy, and consistency as constraint control, which is a multi-dimensional indicator,and it could be express:
\begin{equation}
\mathcal{R}(f) = \alpha_1 \cdot SL(f) + \alpha_2 \cdot PC(f) + \alpha_3 \cdot \log(1+|F_f|) + \alpha_4 \cdot ER(f,h),
\label{control equation}
\end{equation}
where $\alpha_i, i=1,2,3,4$ is the weight coefficients of each term, and $ER(f,h)$ is the comprehensive fitness index of the factor's innovativeness relative to the existing alpha library and its alignment with the investment hypothesis $h$.

\paragraph{Complexity Control}
In Eq.~(\ref{control equation}), $SL(f)$ measures the factor symbolic length, $PC(f)$ counts the number of free parameters and the $\log(1+|F_f|)$ term is used to constrain the excessive use of original features set $|F_f|$ by the factor. The whole three is used to control the factor complexity.

\paragraph{Redundancy Control}
The structural similarity between factors is quantified through AST matching. We capture redundancy through decode $ER(f,h)$ form Eq.~(\ref{control equation}):
\begin{equation}
ER(f,h) = \beta_1 \cdot S(f) + \beta_2 \cdot C(h,d,f),
\label{comprehensive}
\end{equation}
where $\beta_1, \beta_2$ are the weight coefficients of each term, $S(f)$ measures the maximum structural similarity between the factor and the existing factor library to control redundancy. Given two factors $f_i$ and $f_j$ with AST representation $T (f_i)$ and $T (f_j)$, we define $s(f_i, f_j)$ as the size of their largest common isomorphic subtree:
\begin{equation}
s(f_i, f_j) =
\max_{t_i \in T(f_i),\, t_j \in T(f_j)}
\left\{ |t_i| : t_i \cong t_j \right\}
\end{equation}
where $t_i$ and $t_j$ are subtrees of $T(f_i)$ and $T(f_j)$ respectively, $|t_i|$ denotes the size of the subtree (number of nodes), and $t_i$ $t_j$ indicates structural isomorphism between subtrees.With this similarity metric, quantitative measure of $S(f)$ is:
\begin{equation}
S(f) = \max_{f_j \in \mathcal{F}} s(f_i,\phi),
\end{equation}
where $\mathcal{F}$ is the existing alpha factor library and $f_j \in \mathcal{F} = \{f_1, f_2, \dots, f_{i-1}, f_{i+1}, \dots, f_N \}$.

\paragraph{Consistency Control} According to Eq.~(\ref{comprehensive}), $C(h,d,f)$ compute semantic consistency, includes the logical consistency among the factor hypothesis text, factor expression, and code implementation, in order to avoid semantic drift in the LLM generation process, and ensures the interpretability of the factor's economic logic. For a given hypothesis $h$, factor description $d$, and factor expression $f$, we formulate a consistency scoring function:
\begin{equation}
\mathcal{C}(h,d,f) = \alpha \cdot C_1(h,d) + (1-\alpha) \cdot C_2(d,f),
\end{equation}
where the weight coefficient $\alpha$ is set to 0.5; $C_1(h,d)$ is used to evaluate the consistency between $h$ and $d$, $C_2(d,f)$ is used to evaluate the consistency between $d$ and $f$. Factors that fail the verification will trigger regeneration.

\subsubsection{Self-Evolution}
We designs two core evolutionary operators: Mutation and Crossover, which replace the random search mode of traditional GP. This design realizes and diversity guarantee of trajectories, while fully retaining the validated effective investment logic and avoiding semantic drift caused by unconstrained random generation. 

\paragraph{Mutation} Given a mining trajectory $\tau \in \mathcal{T}_{i-1}$, sub-optimal node $a_k$ in $\tau$ is located through self-reflection, then AlphaSeek performs orthogonal mutation on existing trajectories:
\begin{equation}
\tau_{child} = (s_0, a_0, \dots, s_k, \mathrm{Refine}(a_k), s'_{k+1}, a'_{k+1}, \dots, s'_n),
\end{equation}
which only carries out targeted rewriting of local components, historical node $s_k$ and before of $\tau$ are frozen and retaining the core investment logic. It explores independent data dimensions and investment logics on the premise of ensuring logical consistency, thus guaranteeing the diversity of the factor library. 

\paragraph{Crossover} Crossover fuses effective segments of multiple high-return trajectories from different parallel directions, at iteration $i-1$, given $k$ trajectories $\tau^{(1)}, \dots, \tau^{(k)} \in \mathcal{T}_{i-1}$, AlphaSeek recombines them to generate new strategy trajectories:
\begin{equation}
\tau_{\mathrm{child}} = \mathrm{Crossover}(\tau^{(1)}, \dots, \tau^{(k)}).
\end{equation}
which realizing the collaborative amplification of successful experiences from different trajectories. While improving the predictive power of factors, it ensures the interpretability and robustness of factors through explicit self-evolution trajectory.

\subsubsection{Self-Iteraction Portfolio}
Given the factor set $\mathcal{F} = \{f_1, f_2, \ldots, f_M\}$ generated by the trajectory strategy, to extract the optimal combination signals from these static factor pools and construct a robust investment strategy, we design an automated portfolio and backtesting iteration mechanism. Driven by the comprehensive strategy performance, this mechanism conducts continuous exploration and optimization in the portfolio weight space and model configuration space, realizing a full-link closed loop from strategy construction to performance feedback.

The overall cycle consists of three core stages: strategy generation, execution and backtesting, as well as performance feedback and iteration. In the $n$-th round of iteration, the system automatically generates a new strategy construction hypothesis $(\mathbf{w}_n, \Theta_n)$ using the historical record set $\mathcal{D}_{n-1}= \{(\mathbf{w}_i, \Theta_i, \mathbf{s}_i)\}_{i=1}^{n-1}$, where $\mathbf{w}_n$ denotes the portfolio weight vector, and $\Theta_n$ represents the model category and its parameter settings. The strategy hypothesis generation mapping can be expressed as:
\begin{equation}
    (\mathbf{w}_n, \Theta_n) = \mathcal{G}\big(\mathcal{D}_{n-1}\big),
\end{equation}
where $\mathcal{G}(\cdot)$ denotes the strategy hypothesis generation function, whose output is the weight allocation and model configuration of the current round. Based on this hypothesis, the input feature matrix $\mathbf{X}_n\in \mathbb{R}^{T\times M}$ is constructed from the factor pool $\mathcal{F}$, and the corresponding future return label is defined as:
\begin{equation}
    \mathbf{y}_n = \frac{P_{t+1} - P_t}{P_t},
\end{equation}
where $P_t$ represents the closing price of the underlying asset at time $t$.

The execution unit maps the strategy into specific executable tasks, and completes model training and verification on the cross-section of backtesting data. During the execution process, the model architecture category and training process are determined according to the model configuration $\Theta_n$, and preprocessing operations such as cross-sectional normalization are performed on the input matrix $\mathbf{X}_n$ to ensure scale consistency. In the prediction phase, the strategy ranks the prediction scores according to the portfolio weight vector $\mathbf{w}_n$, and constructs a risk-adjusted equivalent return sequence through the Top-k investment strategy and Dropout technology.

The backtesting phase adopts multi-dimensional performance indicators, including Annualized Rate of Return ($ARR$), Information Ratio ($IR$), Maximum Drawdown ($MDD$) and Calmar Ratio ($CR$). These indicators are defined as the comprehensive strategy performance function:
\begin{equation}
    \mathbf{s}_n = \mathcal{S}\big(\mathbf{w}_n, \Theta_n \mid \mathbf{X}_n, \mathbf{y}_n\big),
\end{equation}
where $\mathbf{s}_n = (ARR, IR, MDD, CR)$. To unify the dimension and optimization direction of different indicators, We weight the four indicators $ARR, IR, MDD, CR$ to obtain $R_n$, which not only emphasizes profitability, but also takes into account risk robustness, providing an optimization signal for subsequent iterations.

After completing the strategy construction and evaluation of the current round, the system adds the strategy hypothesis $(\mathbf{w}_n, \Theta_n)$ and its performance $(\mathbf{s}_n, R_n)$ to the historical record set $\mathcal{D}_n$. To achieve an adaptive trade-off between the portfolio weight adjustment direction and the model structure configuration direction, the statistical information of the historical reward sequence $\{U_i\}_{i=1}^{n}$ is introduced to dynamically adjust the direction selection:
\begin{equation}
    d_n = \arg\max_{x \in \{\mathbf{w}_n, \Theta_n\}} \mathbb{E}[U_n \mid x, \mathcal{D}_{n-1}],
\end{equation}
where $d_n$ denotes the optimization direction of the current round. The system dynamically selects the search direction that is more likely to improve the comprehensive performance according to the historical conditional expected value, so as to ensure that the automated cycle can learn from historical experience and adjust the future search strategy in each iteration, realizing the self-iterative optimization of portfolio weights and model parameters.

\section{Experiments}\label{sec5}
\subsection{Experimental Setup}

\subsubsection{Dataset and Backtest}
We conduct experiments on daily frequency data of CSI 300, which covers 300 large-cap A-share stocks in the Chinese market. The dataset is strictly divided into training(Jan.1, 2016 to Dec. 31, 2020), validation(Jan. 1, 2021 to Dec. 31, 2021) and testing(Jan. 1, 2022 to Dec. 26, 2025) according to the time series without look-ahead bias.

The backtesting benchmark is the CSI 300 and it can be extended to zero-shot transfer experiments in markets such as CSI 500.

\subsection{Main Result}\label{subsec2}

Table \ref{main table} and Fig. \ref{Model Comparison} reports the performance of AlphaSeek against a broad set of machine learning models, deep learning models, factor libraries, and recent LLM-based agentic factor mining methods on CSI 300. Overall, AlphaSeek achieves the strongest strategy-level performance among all compared methods, delivering the highest ARR of \textbf{8.28\%}, the best IR of \textbf{1.29}, and the highest CR of \textbf{1.31}, while simultaneously attaining the lowest MDD of \textbf{6.28\%}. These results indicate that AlphaSeek yields a more favorable return–risk trade-off than both conventional predictive models and recent agentic baselines, suggesting that the proposed end-to-end closed loop can translate mined factors into more robust and deployable investment signals rather than merely improving isolated predictive statistics.

\begin{table}[htbp]
  \centering
  \resizebox{\textwidth}{!}{
  \begin{tabular}{llcccccccc}
    \toprule
    \multicolumn{2}{c}{Methods} & \multicolumn{4}{c}{Factor Predictive Power} & \multicolumn{3}{c}{Strategy Performance} \\
    \cmidrule(lr){3-6} \cmidrule(lr){7-10}
    & & IC & ICIR & RIC & RICIR & IR (SHR*) & CR & ARR (\%) & MDD (\%)$\downarrow$ \\
    \midrule
    \multirow{5}{*}{\makecell{Machine \\ Learning}}
    & Linear & 0.0155 & 0.1174 & 0.0368 & 0.2834 & -0.3078 & -0.1407 & -2.67 & 18.97 \\
    & XGBoost & 0.0175 & 0.1336 & 0.0420 & 0.3417 & -0.5280 & -0.1488 & -4.24 & 28.50 \\
    & CatBoost & 0.0162 & 0.1203 & 0.0405 & 0.3289 & -0.2807 & -0.1077 & -2.30 & 21.35 \\
    & LightGBM & 0.0247 & 0.2055 & 0.0423 & 0.3726 & 0.0092 & 0.0032 & 0.07 & 21.80 \\
    & MLP & 0.0321 & 0.2780 & 0.0438 & 0.4088 & 0.1716 & 0.0804 & 1.46 & 18.15 \\
    & DoubleEnsemble & 0.0213 & 0.1670 & 0.0408 & 0.3372 & 0.2490 & 0.1233 & 1.85 & 15.00 \\
    \cmidrule(l){2-10}
    \multirow{4}{*}{\makecell{Deep \\ Learning}}
    & GRU & 0.0321 & 0.2603 & 0.0442 & 0.3601 & 0.5302 & 0.2405 & 3.61 & 15.01 \\
    & Transformer & 0.0331 & 0.2702 & 0.0451 & 0.3801 & 0.4502 & 0.3773 & 5.21 & 13.81 \\
    & LSTM & 0.0331 & 0.2502 & 0.0451 & 0.3503 & 0.6802 & 0.4058 & 6.01 & 14.81 \\
    & TRA & 0.0421 & 0.3402 & 0.0511 & 0.4203 & 1.0502 & 0.8002 & 6.81 & 8.51 \\
    \cmidrule(l){2-10}
    \multirow{3}{*}{\makecell{Factor Libraries}}
    & Alpha158(20) & 0.0051 & 0.0329 & 0.0184 & 0.1177 & 0.5044 & 0.2087 & 4.63 & 22.19 \\
    & Alpha158 & 0.0131 & 0.0817 & 0.0334 & 0.2119 & 0.4099 & 0.2620 & 2.66 & 10.15 \\
    & Alpha360 & 0.0105 & 0.0636 & 0.0306 & 0.1889 & 0.6009 & 0.3550 & 4.09 & 11.52 \\
    \cmidrule(l){2-10}
    \multirow{4}{*}{\makecell{LLM-based \\Agentic\\Factor Mining}}
    & RD-Agent\cite{li2025r} & 0.0401 & 0.3250 & 0.0522 & 0.4250 & 0.8202 & 0.4332 & 7.81 & 18.03 \\
    & AlphaQCM\cite{chen2025alphasage} & 0.0430 & 0.2620 & 0.0420 & 0.2460 & 0.3600 & - & 1.95 & 24.80 \\
    & AlphaForge\cite{chen2025alphasage} & 0.0410 & 0.2590 & 0.0520 & 0.3060 & 0.8800 & - & 3.90 & 21.90 \\
    & \textbf{AlphaSeek(ours)} & \textbf{0.0454} & 0.2561  & 0.0431 & 0.2497 & \textbf{1.29} & \textbf{1.31} & \textbf{8.28} & \textbf{6.28} \\
    \bottomrule
  \end{tabular}
  }
    \caption{Comparison of Factor Predictive Power and Strategy Performance Across Methods}
    \label{main table}
\end{table}

\subsection{Ablation Study}\label{subsec2}
\subsubsection{Ablation of Evolutionary Mining Components}
As shown in Table \ref{ablation 1}, the full model still achieves the best overall results, confirming the effectiveness of integrating these components into a unified trajectory-level evolution pipeline. Removing Planning causes only limited degradation in IC and Rank IC, but leads to a much larger deterioration at the strategy level, with ARR decreasing by \textbf{5.80\%} and MDD increasing by \textbf{4.77\%}, indicating that diversified planning mainly improves the quality of initial search directions and helps discover factors that remain complementary after portfolio construction. Removing Mutation results in the most severe decline in predictive quality, with IC decreasing by \textbf{3.25\%}, alongside ARR decreasing by \textbf{2.46\%} and MDD increasing by \textbf{5.24\%} percentage points, showing that mutation is critical for effective local exploration and trajectory refinement. By contrast, removing Crossover leads to a relatively smaller drop in predictive metrics but a clear deterioration in deployable performance, with ARR decreasing by \textbf{3.55\%} and MDD increasing by \textbf{7.01\%}, suggesting that crossover mainly contributes to robustness by recombining complementary trajectory segments across parallel directions. Overall, the three components play distinct yet complementary roles in expanding search coverage, improving refinement quality, and enhancing strategy robustness.
\begin{table}[htbp]
  \centering
  \caption{Ablation study of evolutionary mining components.}
  \begin{tabular}{>{\raggedright}p{3cm} *{4}{>{\centering\arraybackslash}p{2cm}}}
    \toprule
    \multirow{2}{*}{\makecell[l]{Method}} & \multicolumn{4}{c}{Key Metrics} \\
    \cmidrule(lr){2-5}
    & IC & Rank IC & ARR (\%) & MDD (\%) $\downarrow$ \\
    \midrule
    AlphaSeek & 0.0454 & 0.0437 & 8.28 & 6.28 \\
    \quad - w/o Parallel & 0.0438\rlap{\textcolor{red!60}{$^{-0.0016}$}} & 0.0420\rlap{\textcolor{red!60}{$^{-0.0017}$}} & 2.48\rlap{\textcolor{red!60}{$^{-5.80}$}} & 13.75\rlap{\textcolor{red!60}{$^{+4.77}$}} \\
    \quad - w/o Mutation & 0.0129\rlap{\textcolor{red!60}{$^{-0.0325}$}} & 0.0431\rlap{\textcolor{red!60}{$^{-0.0006}$}} & 5.82\rlap{\textcolor{red!60}{$^{-2.46}$}} & 11.52\rlap{\textcolor{red!60}{$^{+5.24}$}} \\
    \quad - w/o Crossover & 0.0433\rlap{\textcolor{red!60}{$^{-0.0021}$}} & 0.0416\rlap{\textcolor{red!60}{$^{-0.0021}$}} & 4.73\rlap{\textcolor{red!60}{$^{-3.55}$}} & 13.29\rlap{\textcolor{red!60}{$^{+7.01}$}} \\
    \bottomrule
  \end{tabular}
  \label{ablation 1}
\end{table}

\subsubsection{Ablation of Consistency, Complexity, and Redundancy Controls}
\begin{figure}[t]
  \centering
  \includegraphics[width=0.8\linewidth]{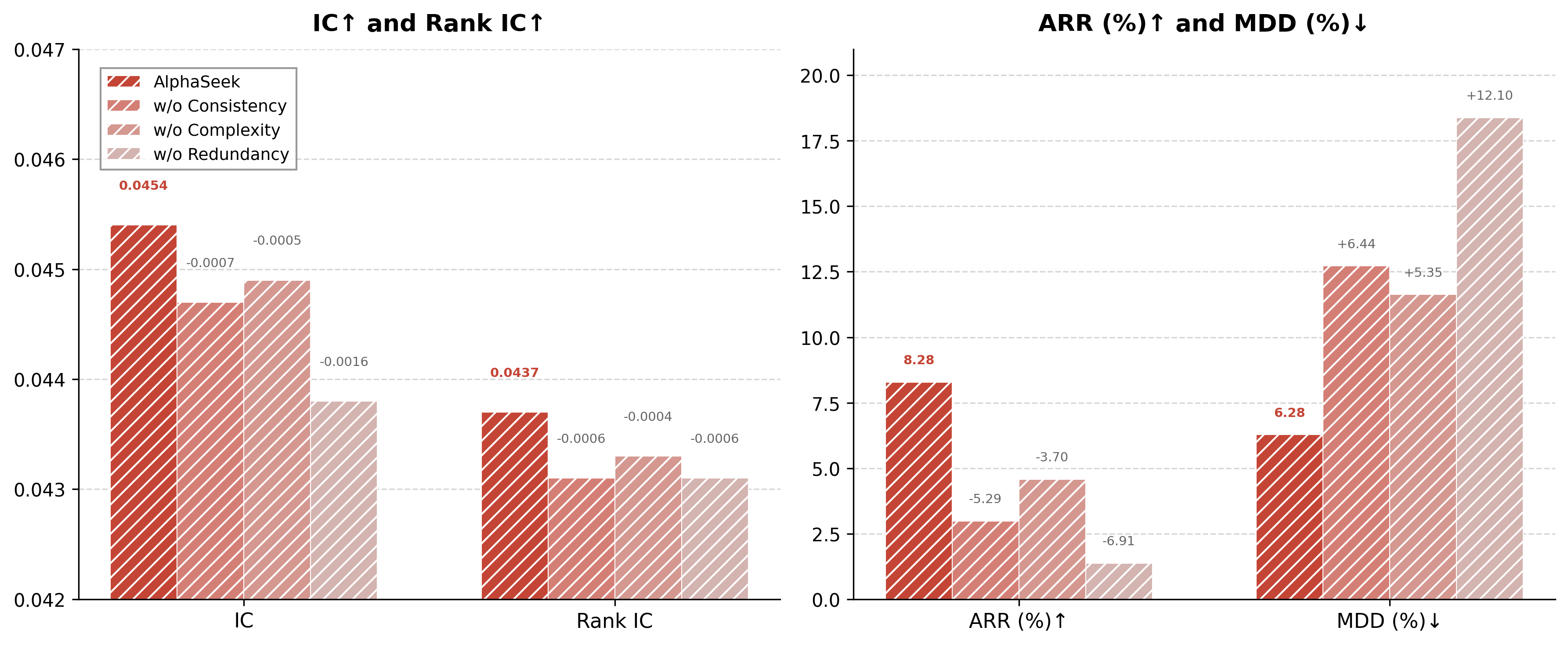}
  \caption{Ablation of Consistency, Complexity, and Redundancy Controls}
  \label{ablation_study_contral}
\end{figure}

As shown in Fig. \ref{ablation_study_contral}, removing any of the three controls leads to a clear performance drop, confirming that they are essential components of AlphaSeek rather than auxiliary design choices. Removing consistency control degrades overall results, indicating that this constraint is important for keeping generated factors economically coherent and executable throughout the mining process. Removing complexity control also weakens performance, but its impact is the smallest among the three, suggesting that it mainly serves to improve robustness by discouraging overly intricate expressions. In contrast, removing redundancy control causes the most pronounced deterioration, including the largest drop in IC, ARR decreasing by \textbf{6.91\%}, and MDD increasing by \textbf{12.10\%}, showing that redundancy filtering is especially important for preserving factor diversity and preventing ineffective concentration in the final factor pool. Overall, the three controls play complementary roles in maintaining factor validity, robustness, and diversity.

\subsubsection{Ablation of Self-Iteraction Portfolio}
Fig. \ref{Ablation self_Iteraction} shows that AlphaSeek consistently dominates the variant without self-iteration across IC, Rank IC, ARR, and inverse MDD, resulting in a substantially larger radar area overall. The largest gain appears on ARR, indicating that self-iteration is particularly important for converting mined factors into effective portfolio returns.

\subsection{More Analysis}
\subsubsection{Alpha Mining Efficiency Analysis}\label{subsubsec2}
\begin{figure}[t]
  \centering
  \begin{minipage}{0.42\linewidth}
    \centering
    \includegraphics[width=\linewidth]{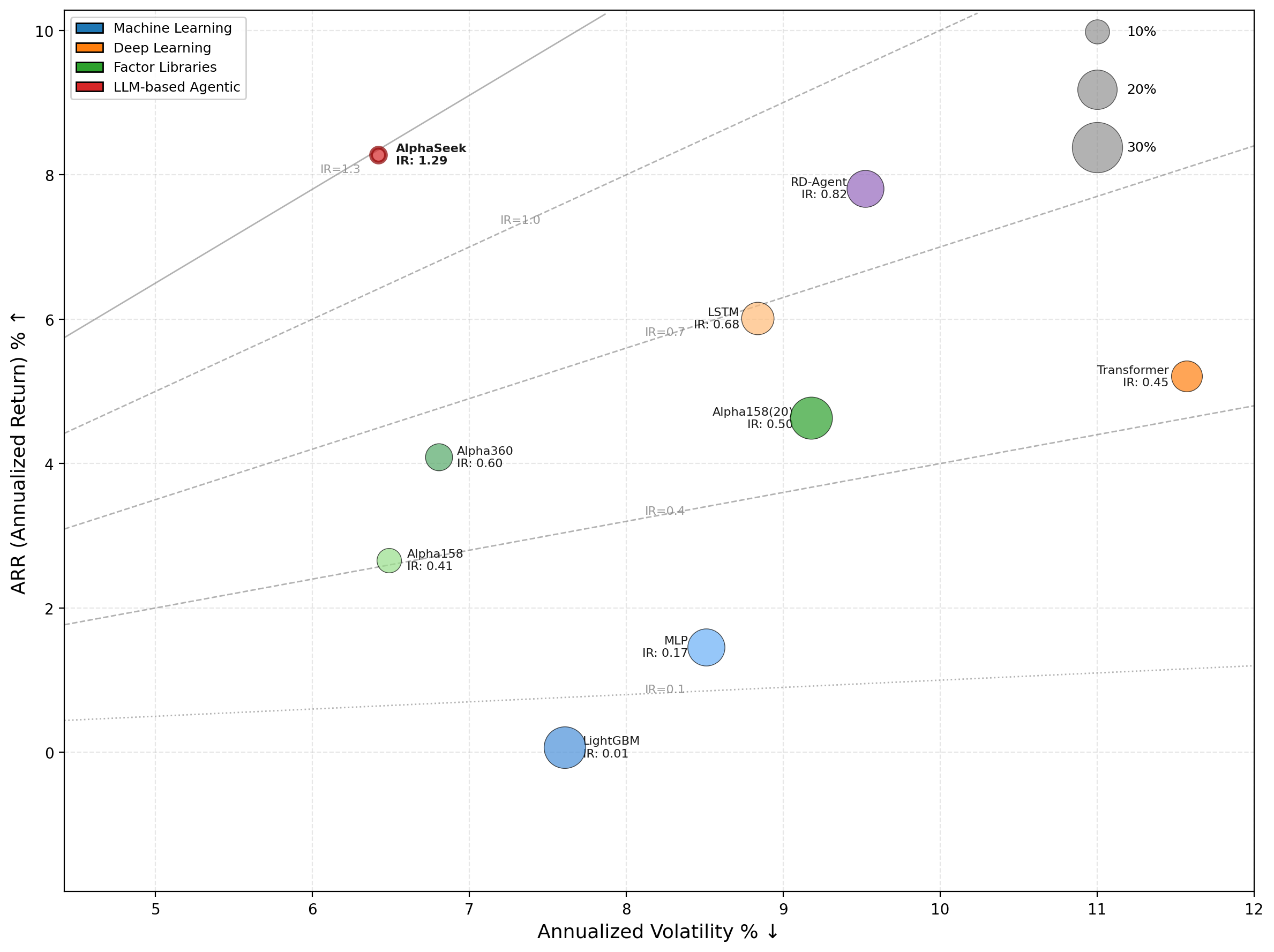}
    \caption{Model Performance: AER vs volatility with MDD and IR}
    \label{fig:ARE_vs_AV}
  \end{minipage}
  \hspace{0.02\linewidth}
  \begin{minipage}{0.42\linewidth}
    \centering
    \includegraphics[width=\linewidth]{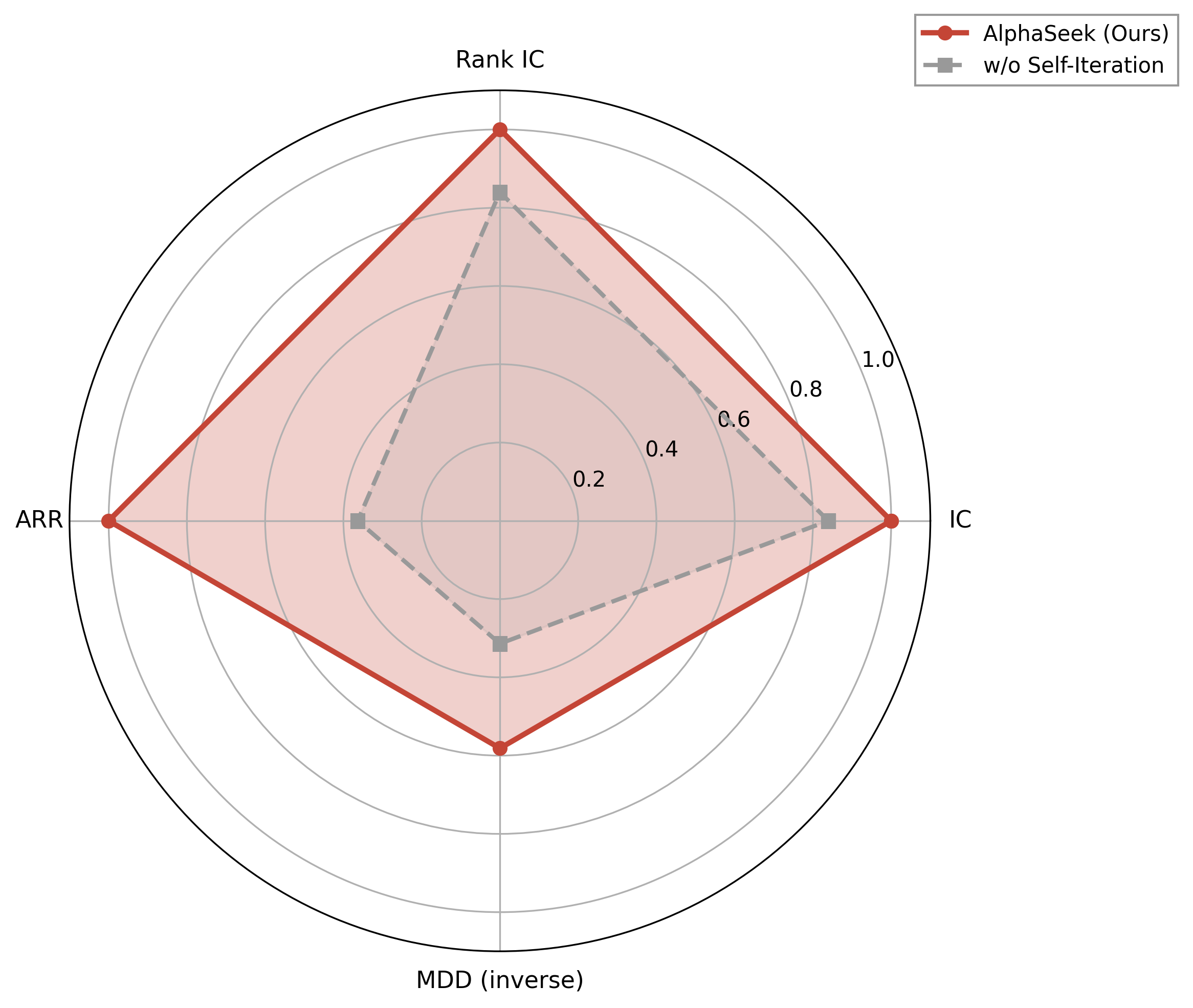}
    \caption{Ablation self-Iteraction}
    \label{Ablation self_Iteraction}
  \end{minipage}
\end{figure}

Following the risk–return analysis, Fig. \ref{fig:ARE_vs_AV} compares different methods in the AER–volatility space, while additionally encoding MDD through marker size and IR through iso-performance lines. AlphaSeek is located near the upper-left frontier of the plot, indicating that it achieves a more favorable balance between excess return and risk exposure than competing methods. In particular, compared with prior LLM-based agents, AlphaSeek attains higher annualized excess return under lower volatility, while also exhibiting a visibly smaller drawdown footprint, which suggests that its advantage does not come from simply taking more aggressive risk. Instead, our mechanism improves the efficiency with which predictive signals are converted into deployable strategy returns.

\subsubsection{Time-Series Returns on CSI500}\label{subsubsec2}
Fig. \ref{fig:csi500} shows the time-series returns on CSI 500, where all factors are mined on CSI 300 and transferred without any market-specific re-optimization. AlphaSeek maintains the strongest return trajectory and finishes above all competing methods, demonstrating that the discovered factors remain effective under a clear shift in market universe. This result suggests that AlphaSeek captures more transferable return structures rather than dataset-specific patterns. Moreover, its return curve exhibits stronger resilience during market fluctuations and a more sustained upward trend in later periods, while other methods tend to stagnate or lag behind after intermediate volatility.

\section{Conclusion}

Overall, we propose AlphaSeek, an end-to-end LLM-driven Alpha mining agent aiming at the problems of subjective direction design, insufficient multi-source integration, semantic drift, and lack of a complete end-to-end factor discovery pipeline. Specifically, through the trajectory-level factor evolution framework, the agent upgrades the optimization unit from a single factor to the complete research life cycle, realizing the intergenerational inheritance and interpretability of factor knowledge. Through the self-iteraction portfolio mechanism, it constructs a full-process closed loop of mining-portfolio-backtesting. Empirical results based on the CSI 300 market show that the method proposed in this paper overall outperforms mainstream benchmark methods with ARR of \textbf{8.28\%}, IR of \textbf{1.29\%} and MDD of \textbf{6.28\%}, while remaining competitive on factor predictive metrics with IC of \textbf{0.0454}, and achieve strong time-series return performance in out-of-sample darases CSI500.

\newpage

%

%
\bibliographystyle{splncs04}
\bibliography{mybibliography}

\end{document}